\documentclass[aps,prd,twocolumn,preprintnumbers,amsmath,amssymb,
superscriptaddress,showpacs,floatfix]{revtex4}

\usepackage{graphicx}
\usepackage{dcolumn}
\usepackage{bm}
\usepackage{epsfig}
\usepackage{graphics}
\usepackage{latexsym}

\def\beq{\begin{equation}}
\def\eeq{\end{equation}}
\def\bea{\begin{eqnarray}}
\def\eea{\end{eqnarray}}
\newcommand{\beqs}{\begin{subequations}}
\newcommand{\eeqs}{\end{subequations}}

\newcommand{\cref}[1]{Ref.~\cite{#1}}

\newcommand{\vev}[1]{\left<#1\right>}

\newcommand{\hh}{{\ensuremath{I{\kern-2.6pt h}}}}
\newcommand{\bhh}{{\ensuremath{\bar{I{\kern-2.6pt h}}}}}

\begin{document}

\preprint{UT-STPD-16/02}

\title{Axion, $\mu$ Term, and Supersymmetric Hybrid Inflation}

\author{G. Lazarides}
\email{lazaride@eng.auth.gr} \affiliation{School of Electrical and
Computer Engineering, Faculty of Engineering, Aristotle University
of Thessaloniki, Thessaloniki 54124, Greece}
\author{C. Panagiotakopoulos}
\email{costapan@eng.auth.gr}
\affiliation{School of Rural and Surveying Engineering, Faculty of 
Engineering, Aristotle University of Thessaloniki, Thessaloniki 
54124, Greece}
\author{Q. Shafi}
\email{shafi@bartol.udel.edu}
\affiliation{Bartol Research Institute, Department of Physics and 
Astronomy, University of Delaware, Newark, DE 19716, USA}

\date{\today}

\begin{abstract}
We show how successful supersymmetric hybrid inflation is realized in 
realistic models where the resolution of the minimal supersymmetric 
standard model $\mu$ problem is intimately linked with axion physics. 
The scalar fields that accompany the axion, such as the saxion, are 
closely monitored during and after inflation to ensure that the axion 
isocurvature perturbations lie below the observational limits. The 
scalar spectral index $n_{\rm s}\simeq 0.96 - 0.97$, while the 
tensor-to-scalar ratio $r$, a canonical measure of gravity waves, 
lies well below the observable range in our example. The axion domain 
walls are inflated away, and 
depending on the axion decay constant $f_{\rm a}$ and the magnitude 
of the $\mu$ parameter, the axions and/or the lightest supersymmetric 
particle compose the dark matter in the universe. Non-thermal 
leptogenesis is naturally implemented in this class of models.
\end{abstract}

%\keywords{MSSM,
%Hyperbolic Branch, Focus Point, Higgs Boson, Dark Matter, 
%Neutralino, Higgsino}
%Use showkeys class option if keyword display desired
\maketitle

\section{Introduction}

An attractive feature of the simplest supersymmetric (SUSY) hybrid 
inflation models \cite{dvali,copeland} (for a review see e.g., 
Ref.~\cite{lectures}) is the prediction that, to a good 
approximation, the temperature anisotropy $\delta T/T$ of the 
cosmic microwave background radiation is proportional to 
$(M/m_{\rm P})^2$ \cite{dvali}, where $M$ denotes the scale of the 
gauge symmetry breaking associated with inflation and $m_{\rm P}
\simeq 2.44\times 10^{18}~{\rm GeV}$ denotes the reduced Planck mass. 
Thus, $M$ is comparable to or somewhat smaller than the SUSY grand 
unified theory (GUT) scale $M_{\rm GUT}$. Moreover, the scalar 
spectral index $n_{\rm s}$ lies in the observed range of 0.96-0.97 
\cite{planckcosm} provided the inflationary potential incorporates, 
in addition to radiative corrections \cite{dvali}, either the soft 
SUSY breaking terms \cite{rehman,gravitywaves} or higher order 
terms in the K\"{a}hler potential \cite{bastero}. In the absence 
of these terms the spectral index $n_{\rm s}$ lies close to 0.98, 
which seems acceptable provided the effective number of light 
neutrino species is slightly higher than 3 \cite{planckcosm}. 
Note that in either case the supergravity corrections 
remain under control because the 55 or so e-foldings necessary 
to resolve the horizon and flatness problems are realized for 
inflaton values well below the Planck scale $m_{\rm P}$. Finally,
the initial condition problem of hybrid inflation can be solved 
\cite{initial} by invoking two stages of inflation. 

In this paper, we propose to merge SUSY hybrid inflation with 
axion physics, which provides \cite{PecceiQuinn} the most 
compelling resolution of the strong CP problem. We incorporate 
in our 
discussion the nice feature that the intermediate ($M_{\rm I}
\sim 10^{11}-10^{12}~{\rm GeV}$ \cite{preskill}) spontaneous 
symmetry breaking scale of the global $U(1)$ axion symmetry 
can be exploited \cite{nilles} to resolve the $\mu$-term 
problem of the minimal supersymmetric standard model (MSSM). 
Namely, the usual MSSM $\mu$ term is forbidden through a 
combination of the underlying symmetries, but it arises from a 
higher dimensional superpotential term. The resulting value of 
$\mu $ turns 
out to be of order $M_{\rm I}^2/m_{\rm P}\sim {\rm TeV}$, the 
desired magnitude. The scalar spectral index $n_{\rm s}$ is 
adjusted to the observed range by using higher order terms in 
the K\"{a}hler potential. The tensor-to-scalar ratio $r$ turns 
out to be negligible in this simple hybrid inflation model. 
Appreciable values of $r$ could, however, be achieved if one 
replaces this model by a two stage hybrid inflation scenario 
\cite{dinf}.    

Note that this extension of the MSSM with matter parity 
allows for the possible presence of more than a single 
species of cold dark matter. With a suitable choice of the 
parameters both the axion and the lightest supersymmetric 
particle (LSP) would contribute to the dark matter density 
in the universe. For a recent discussion of natural SUSY in 
the framework of axion physics see Ref.~\cite{baer} and 
references therein.

In particular, in this work, we investigate how the 
Peccei-Quinn (PQ) or axion transition proceeds in a SUSY 
model with hybrid inflation without creating any problems 
with the observations. A potential problem is the generation 
of unacceptably large axion isocurvature 
perturbations. The avoidance of such perturbations requires 
that the value of the PQ field during inflation be kept 
fairly large, which can be achieved by appropriately using 
higher order terms in the K\"{a}hler potential. These terms 
produce a suitable negative mass-squared term for the PQ 
field and thereby shift its vacuum value to become 
appropriately large during inflation. Consequently, the 
value of the PQ field remains large during inflation. It 
also remains large but time-dependent during the inflaton 
oscillations which follow inflation and gradually
drifts to the desired low energy PQ vacuum. As a consequence,
the PQ symmetry is spontaneously broken already during 
inflation and remains so thereafter. Therefore, the axion 
domain wall problem \cite{sikivie}, which appears if the 
spontaneous breaking of the PQ symmetry takes place after 
inflation, does not arise here. For a different approach to
avoid domain walls in axion models see Ref.~\cite{yanagida}.     

A thorough numerical investigation over several orders of 
magnitude in cosmic time shows how the PQ field value 
develops during inflation and the subsequent inflaton 
oscillations until low energies where the soft SUSY 
breaking terms take over. We estimate 
the amplitude of the PQ oscillations when this regime is 
reached in order to calculate the ensuing relic saxion 
abundance. Note that the system contains not only the 
saxion and the axion but also an extra complex Higgs 
field and a four component Dirac axino. Under certain 
conditions, one can show that the extra Higgs field does 
not perform any appreciable oscillations which could 
contribute to the energy density of the universe. 

The decay time of the various states of the PQ system and the 
relic density of the thermally produced saxions, axinos, and 
extra Higgs particles have been estimated too. The possible 
contribution of the axinos as well as the axions to 
the cold dark matter of the universe is briefly explored.

The layout of the paper is as follows. In Sec.~\ref{trans}, 
we investigate the PQ transition in the context of SUSY hybrid
inflation. In Sec.~\ref{relic}, we estimate the relic density
of the various PQ states. Sec.~\ref{sec:iso} is devoted to the 
discussion of the axion isocurvature perturbations and our 
conclusions are summarized in Sec.~\ref{concl}.   

\section{Model and the PQ transition}
\label{trans}

For definiteness, we consider a SUSY model based on the 
left-right symmetric gauge group $G_{\rm{LR}}=SU(3)_{\rm c}
\times SU(2)_{\rm L}\times SU(2)_{\rm R}\times U(1)_{B-L}$. 
The superfield content of the model comprises the gauge 
singlets $S$, $N$, $\bar N$, the Higgs superfields $H$, 
$\bar{H}$, $h$ belonging to the $(1,1,2)_{1}$, $(1,1,2)_{-1}$, 
$(1,2,2)_{0}$ representation of $G_{\rm{ LR}}$, respectively,
and the matter superfields $q_i$, $q^c_i$, $l_i$, $l^c_i$ 
(with $i=1,2,3$) which belong to the $(3,2,1)_{\frac{1}{3}}$, 
$(\bar 3,1,2)_{-\frac{1}{3}}$, $(1,2,1)_{-1}$,
$(1,1,2)_{1}$ representation of $G_{\rm LR}$, respectively.
Here the subscripts denote $U(1)_{B-L}$ charges.
The fields  $H$ and $\bar{H}$ acquire vacuum expectation 
values (VEVs), which cause the breaking of $G_{\rm LR}$ to the 
standard model (SM) gauge group $G_{\rm SM}$, while the VEV 
of $h$ is responsible for the electroweak symmetry breaking.

The global symmetries imposed on the model include a R 
symmetry $U(1)_{\rm R}$ under which the superfields have the 
following charge assignments
\begin{eqnarray}
R:& & S(2), N(0), \bar N(2), H(0), \bar H(0), h(1),
\nonumber\\
& & q_i(0), q^c_i(1), l_i(0), l^c_i(1)
\end{eqnarray}
with the superpotential having charge 2, and a PQ symmetry 
$U(1)_{\rm {PQ}}$ under which the superfields have the 
following charges
\begin{eqnarray}
PQ:& & S(0), N(-1), \bar N(3), H(0), \bar H(0), h(1),
\nonumber\\
& & q_i(-1), q^c_i(0), l_i(-1), l^c_i(0).
\end{eqnarray}
The above charge assignments allow in the superpotential the 
usual Yukawa couplings for quarks and leptons as well as the 
higher order terms $\bar{H}\bar{H}l^c_i l^c_j$ which 
provide intermediate scale right-handed neutrino masses. 
Moreover, the MSSM $\mu$ term $h^2$ is forbidden, but the 
term $N^2h^2$ is allowed and generates, as we will see, a 
$\mu$ term of the desired magnitude.

The soft SUSY breaking terms break the $U(1)_{\rm R}$ 
symmetry explicitly to its $Z_2$ subgroup under which 
$h$, $q^c_i$, and $l^c_i$ change sign. This symmetry, 
combined with the $Z_2$ center of $SU(2)_{\rm L}$ under 
which  $h$, $q_i$, and $l_i$ flip sign, provides the 
well-known $Z_2^{\rm mp}$ matter parity symmetry under 
which the matter fields $q_i$, $q^c_i$, $l_i$, and 
$l^c_i$ are odd. Consequently, the apparent spontaneous 
breaking by the VEV of $h$ of the $Z_2$ subgroup of 
$U(1)_{\rm R}$ which is left unbroken after soft SUSY 
breaking does not lead to the cosmologically 
unacceptable production of domain walls. In reality, 
this $Z_2$ symmetry is merely 
replaced by the equivalent $Z_2^{\rm mp}$, which remains 
unbroken. In addition, the $U(1)_{\rm PQ}$ symmetry is 
broken explicitly by QCD instantons to its $Z_6$ subgroup. 
However, the spontaneous breaking of this discrete 
symmetry does not cause any trouble with domain walls. 
The reason is that, as we will demonstrate in the 
following, this breaking occurs already during inflation 
and the walls are inflated away.

It is worth noting that, as a consequence of the field 
content and the symmetries imposed, our simple model
exhibits, excluding non-perturbative effects, exact 
baryon number conservation as an accidental symmetry.  

The superpotential component which is relevant for inflation
and the evolution of the PQ system is
\begin{equation}
W=\kappa S\left(M^2-H{\bar H}\right)+\frac{\lambda}{ m_{\rm P}} 
N^3 \bar N,
\label{W}
\end{equation}
and the K\"{a}hler potential is taken to be
\begin{eqnarray}
\label{kaehler}
&&K=|S|^2+|H|^2+|\bar H|^2+|N|^2+|\bar N|^2 +\frac{\alpha}
{4m_{\rm P}^{2}}|S|^4+\nonumber \\
&&\left( c_1|S|^2+c_2| H|^2+c_3|\bar H|^2+ c_4 
\left(H\bar H+h.c.\right)\right)\frac{|N|^2}{m_{\rm P}^{2}}+
\nonumber\\
&&\left( c_5|S|^2+c_6| H|^2+c_7|\bar H|^2+ c_8 
\left(H\bar H+h.c.\right)\right)\frac{|\bar N|^2}{m_{\rm P}^{2}}.
\nonumber\\
\end{eqnarray}
Here $M$ is a superheavy mass and $\kappa$, $\lambda$, $\alpha$, 
$c_1$, $c_2$, $c_3$, $c_4$, $c_5$, $c_6$, $c_7$, $c_8$ are 
dimensionless constants which are all taken to be real. In 
addition, we assume that $\alpha>0$ and $c_1>1$, $c_2$, $c_3\ge 1$, 
while $c_5$, $c_6$, $c_7<1$. The motivation for this assumption 
will become clear later. Note that the superpotential in 
Eq.~(\ref{W}) was first introduced in Ref.~\cite{intermediate} 
for achieving the PQ transition after the termination of 
inflation without runaway problems. 

Expanding the F-term potential in powers of $ m_{\rm P}^{-1}$ 
and keeping terms up to $ m_{\rm P}^{-2}$, we obtain
\begin{eqnarray}
\label{pot1}
V&=&\kappa^2|M^2-H \bar H|^2\left(1+|H|^2+|\bar H|^2-
\alpha|S|^2-\right.\nonumber\\
& &\left.(c_1-1)|N|^2+(1-c_5)|\bar N|^2\right)+\nonumber\\
& &\kappa^2|S|^2\left(|H|^2+|\bar H|^2\right)\left(1+|S|^2
+|H|^2+|\bar H|^2\right)+\nonumber\\
& &4\kappa^2|S|^2|H|^2|\bar H|^2-2\kappa^2M^2|S|^2
\left(H\bar H+h.c.\right)-\nonumber\\
& &\kappa^2|S|^2\left((c_3-1)|H|^2+ (c_2-1)
|\bar H|^2\right)|N|^2+\nonumber\\
& &\kappa^2|S|^2\left((1-c_7)|H|^2+ (1-c_6)|\bar H|^2\right)
|\bar N|^2+\nonumber\\
& &\lambda^2|N|^4\left(|N|^2+9|\bar N|^2\right)+\nonumber\\
& &m_{{3}/{2}}^2\left(\varepsilon_1|N|^2+\varepsilon_2
|\bar N|^2\right)
+m_{{3}/{2}}\left(\lambda A N^3 \bar N+h.c.\right).\nonumber\\
\end{eqnarray}
Here we have included the soft SUSY breaking terms associated 
only with the fields $N$ and $\bar N$ and have chosen units in 
which 
the reduced Planck mass scale $m_{\rm P}=1$, which we will use 
in the rest of the paper. Notice that in this approximation 
the parameters $c_4$, $c_8$ do not enter the potential.

For simplicity, in the following we set $A=0$ for the parameter 
of the trilinear soft SUSY breaking terms. We also take 
$\varepsilon_2>0$ for the soft SUSY breaking mass-squared term 
associated with the field $\bar N$. Then, it is apparent that, 
with our assumption that $c_5$, $c_6$, $c_7<1$, $\bar N$ enters 
the potential through a positive definite quadratic term and, 
as a consequence, the minimum of the potential is obtained for 
$\bar N=0$ independent of the values of the remaining fields. 
Therefore, we may safely set $\bar N=0$ throughout the rest of 
our discussion, which considerably simplifies our analysis. 
With $\bar N$ set to 
zero, the parameter $\varepsilon_1$ may be absorbed into a 
redefinition of $m_{{3}/{2}}^2$ once the choice of its sign is 
made. In order to achieve spontaneous breaking of the PQ 
symmetry, we choose $\varepsilon_1<0$. Moreover, the SUSY minimum
of the potential (obtained in the limit $m_{{3}/{2}}\rightarrow 
0$) lies along the D-flat direction $H=\bar H^*$ to which we 
restrict our analysis. 

Using the $U(1)_{\rm R}$,  $U(1)_{\rm {PQ}}$, and $U(1)_{B-L}$
symmetries of the model, we can rotate $S$, $N$, $H$, and, on 
account of the D-flatness, $\bar H$ on the real axis. Then, 
provided that the values of the fields remain well below unity 
in size, we define almost canonically normalized real scalar 
fields $\sigma$, $\chi$, and $h$ as follows 
\begin{equation}
S=\frac{\sigma}{\sqrt{2}}, \quad N=\frac{\chi}{\sqrt{2}}, 
\quad H=\bar{H}=\frac{h}{2}.
\end{equation}
In terms of the fields $\sigma$, $\chi$, and $h$, the potential
becomes
\begin{eqnarray}
\label{pot2}
V&=&\kappa^2 \left(M^2-\frac{h^2}{4}\right)^2
 \left(1+\frac{h^2}{2}-\alpha \frac{\sigma^2}{2}
- d_1 \frac{\chi^2}{2}\right)+\nonumber\\
& &\frac{\kappa^2}{4}\sigma^2h^2\left(
1+\frac{\sigma^2}{2}+h^2-2M^2-d_2\frac{\chi^2}{2}\right)
+\nonumber\\
& &\frac{\lambda^2}{8}\chi^6-\frac{1}{2}m_{{3}/{2}}^2\chi^2,
\end{eqnarray}
where 
\begin{equation}
d_1\equiv c_1-1>0
\end{equation} 
and 
\begin{equation}
d_2\equiv \frac{c_2+c_3}{2}-1\ge 0.
\end{equation}

The part of the potential involving the field $\chi$  
may be written as
\begin{equation}
V_{\chi}=-\frac{1}{2}m_{\chi}^2 \chi^2+
\frac{\lambda^2}{8}\chi^6,
\end{equation}
where
\begin{equation}
m_{\chi}^2 = d_1\,\kappa^2 \left(M^2-\frac{h^2}{4}\right)^2
+d_2\,\frac{\kappa^2}{4}\sigma^2h^2+m_{{3}/{2}}^2.
\label{mchi2}
\end{equation}
$V_\chi$ is minimized for values $\chi_{\rm m}$ 
of $\chi$ satisfying
\begin{equation}
|\chi_{\rm m}|=\sqrt{2}\left(\frac{m_{\chi}^2}
{3\lambda^2}\right)^{\frac{1}{4}}.
\label{potmin}
\end{equation}
Note that the positivity of $d_1$, and non-negativity of 
$d_2$, which are consequences of our assumption that 
$c_1>1$, $c_2$, $c_3\ge 1$, guarantee the positivity of 
$m_{\chi}^2$ in Eq.~(\ref{mchi2}). As a result, during 
inflation, the subsequent inflaton oscillations, and 
after reheating, the minimum of the potential is shifted 
away from $\chi=0$. This is crucial for controlling the 
axion isocurvature perturbations and avoiding the axion 
domain wall problem (see Sec.~\ref{sec:iso}).    

If $\sigma$ satisfies the inequality $\sigma^2>
\sigma_{\rm c}^2\simeq 2M^2$, the $\sigma$-dependent mass 
squared  $m_{ h}^2\simeq\kappa^2(\sigma^2-2M^2)/2$ of $h$ 
is positive, thus, rendering the choice $h=0$ stable. With 
$h=0$ the potential becomes
\begin{equation}
V\overset{h=0}=\kappa^2M^4\left(1-\alpha \frac{\sigma^2}{2}
\right)+V_{\chi}\simeq  \kappa^2M^4,
\end{equation}
leading to a hybrid inflationary stage. In addition, we 
have
\begin{equation}
m_{\chi}^2 = d_1\kappa^2 M^4+m_{{3}/{2}}^2 \simeq 
d_1\kappa^2 M^4
\end{equation} 
and, as a consequence, $|\chi_{\rm m}|$ during inflation 
becomes
\begin{equation}
|\chi_{\rm inf}|\simeq \sqrt{2}\left(\frac{ d_1
\kappa^2 M^4}{3\lambda^2}\right)^{\frac{1}{4}}.
\label{chiinf}
\end{equation}
Since $m_{\chi}$ is of the order of the inflationary 
Hubble parameter, $\chi$ is expected to quickly reach
the above value. With such a value of $\chi$ it can be 
verified that $V_{\chi} \ll  \kappa^2M^4$ and the 
approximation made is justified.

To the inflationary potential we add the contribution
\begin{equation}
V_{\rm {rad}}= \kappa^2M^4\left(\frac{\delta_{h}}{2}\right)
\ln \frac{\sigma^2}{\sigma_{\rm c}^2}
\end{equation}
from radiative corrections, where
\begin{equation}
\delta_{h}=N_h\frac{\kappa^2}{8\pi^2}
\label{deltah}
\end{equation}
and $N_h=2$ is the dimensionality of the representation to 
which $H$, $\bar H$ belong. Note that this equation holds 
for $\sigma/\sigma_{\rm c}\gg 1$, which requires that 
$|\kappa|$ be not much smaller than about $0.01$. The 
potential during inflation is then 
\begin{equation}
V_{\rm {inf}}\simeq \kappa^2M^4\left(1+\frac{\delta_{h}}
{2}\ln \frac{\sigma^2}{\sigma_{\rm c}^2}
-\alpha \frac{\sigma^2}{2}\right)\simeq  \kappa^2M^4.
\end{equation}

The first, second, and third derivative of $V_{\rm inf}$  with 
respect to the inflaton field $\sigma$ are, respectively, given
by
\begin{equation}
V_{\rm {inf}}^{\prime}=\kappa^2M^4\left(
\frac{\delta_{h}}{\sigma}\right)\left(
1-\frac{\alpha\sigma^2}{\delta_h}\right),
\end{equation}
\begin{equation}
V_{\rm {inf}}^{\prime\prime}=- \kappa^2M^4\left(
\frac{\delta_{h}}{\sigma^2}\right)
\left(1+\frac{\alpha\sigma^2}{\delta_h}\right),
\end{equation}
and
\begin{equation}
V_{\rm {inf}}^{\prime\prime\prime}=2 \kappa^2M^4
\frac{\delta_{h}}{\sigma^3}.
\end{equation}
Introducing the parameter $x$
\begin{equation}
x \equiv \frac{\alpha\sigma^2}{\delta_h}<1,
\end{equation}
we obtain the slow-roll parameters
\begin{equation}
\epsilon\equiv\frac{1}{2}\left(\frac{V_{\rm inf}^{\prime}}
{V_{\rm inf}}\right)^2
\simeq \frac{1}{2}\alpha \delta_h\frac{(1-x)^2}{x},
\end{equation}
\begin{equation}
\eta\equiv\frac{V_{\rm {inf}}^{\prime \prime}}{V_{\rm inf}}
\simeq -\alpha \frac{1+x}{x},
\end{equation}
and
\begin{equation}
\xi\equiv\left(\frac{V_{\rm {inf}}^{\prime}}{V_{\rm inf}}
\right)\left(\frac{V_{\rm {inf}}^{\prime \prime\prime}}
{V_{\rm {inf}}}\right)\simeq 2\alpha^2 \frac{1-x}{x^2}.
\end{equation}
From these equations, we see that $\epsilon/|\eta|<\delta_h/2$ 
and $\epsilon |\eta|/\xi <\delta_h/4$. So assuming $\delta_h 
\ll 1$, we obtain $\epsilon \ll |\eta|$ and $\epsilon |\eta| 
\ll \xi$.

Inflation ends when $\sigma$ reaches the value 
$\sigma_{\rm {end}}$ with
\begin{equation}
\sigma_{\rm {end}}^2\simeq \max\{2M^2, \delta_h \},
\end{equation}
depending on whether termination of inflation occurs through 
the waterfall mechanism or because of the radiative 
corrections becoming strong ($|\eta| \simeq 1$). The number 
${\bf N}(x_{\rm {in}})$ 
of e-foldings in the slow-roll approximation for the period 
in which $\sigma$ varies between an initial value 
$\sigma_{\rm {in}}$ and the final value $\sigma_{\rm {end}}$ 
corresponding, respectively, to the values $x_{\rm {in}}$ 
and
\begin{equation}
x_{\rm {end}}\simeq\alpha \max\{ \frac{2 M^2}{\delta_h}, 1\}
\label{xend}
\end{equation}
of the variable $x$ is given by
\begin{equation}
{\bf N}(x_{\rm {in}})\simeq\frac{1}{2\alpha}\ln 
\frac{1-x_{\rm {end}}}{1-x_{\rm {in}}}.
\end{equation}

Let the value of the inflaton field $\sigma$ at horizon exit 
of the pivot scale $k_*=0.05 \ \rm {Mpc}^{-1}$ be $\sigma_*$, 
with $x_*$ being the corresponding value of the parameter $x$. 
Let us also assume that
\begin{equation}
x_{\rm {end}} \ll x_*.
\end{equation}
Then,
\begin{equation}
{\bf N}_* \equiv {\bf N}(x_*) \simeq \frac{1}{2\alpha}\ln 
\frac{1}{1-x_*},
\label{N*}
\end{equation}
which implies that
\begin{equation}
\alpha \simeq \frac{1}{2{\bf N}_*}\ln \frac{1}{1-x_*}.
\end{equation}
Moreover, the scalar spectral index $n_{\rm s}$ is obtained 
in terms of $\eta_*$, the slow-roll parameter $\eta$ 
evaluated at the value $x=x_*$, as follows:
\begin{eqnarray}
n_{\rm s}& \simeq& 1+2\eta_* \nonumber\\
&\simeq& 1-2\alpha\frac{1+x_*}{x_*}\nonumber\\
&\simeq& 1-\frac{1}{{\bf N}_*}
\left(\frac{1+x_*}{x_*}\ln \frac{1}{1-x_*}\right).
\end{eqnarray}
Its running $\alpha_{\rm s}\equiv d n_{\rm s}/d \ln{k}$ is
\begin{equation}
\alpha_{\rm s}\simeq -2\xi_* \simeq -4\alpha^2 
\frac{1-x_*}{x_*^2},
\end{equation}
where $\xi_*$ is the parameter $\xi$ evaluated 
at horizon exit of the pivot scale. Finally, the scalar 
potential on the inflationary path can be written in terms 
of the scalar power spectrum amplitude $A_{\rm s}$ and the 
value $\epsilon_*$ of the slow-roll parameter $\epsilon$, 
both evaluated at horizon exit of the pivot scale, as
\begin{equation}
V=24\pi^2\epsilon_*A_{\rm s},
\end{equation}
from which we obtain
\begin{equation}
M^4\simeq 3\alpha\left(\frac{N_h}{2}\right)
\frac{(1-x_*)^2}{x_*}A_{\rm s}.
\end{equation}

Many SUSY inflation scenarios predict the relation $n_{\rm s}
\simeq 1-1/{\bf N}_*$ (see e.g., Ref.~\cite{dvali}), which gives 
$n_{\rm s}$ values close to 0.98, instead of the presently 
favored values which are close to 0.96 \cite{planckcosm}.
For three neutrino species, a modified relation $n_{\rm s} 
\simeq 1-2/{\bf N}_*$ is in better agreement with recent 
observations. Such a relation is easily obtained in our model 
if we choose for $x_*$ a value close to 0.5. Indeed, setting 
$x_*=0.5$ and ${\bf N}_*=52$, for solving the horizon and 
flatness problems for reheat temperature $T_{\rm r}\sim 
10^9~{\rm GeV}$, we obtain $\alpha\simeq 1/150$, $n_{\rm s}
\simeq 0.96$, $\alpha_{\rm s} \simeq -3.56 \times 10^{-4}$, and 
$M \simeq 2.17 \times 10^{-3}$, setting $N_h=2$ and 
$A_{\rm s}=2.215 \times 10^{-9}$ \cite{planckcosm}.

\begin{figure}[t]
\centerline{\epsfig{file=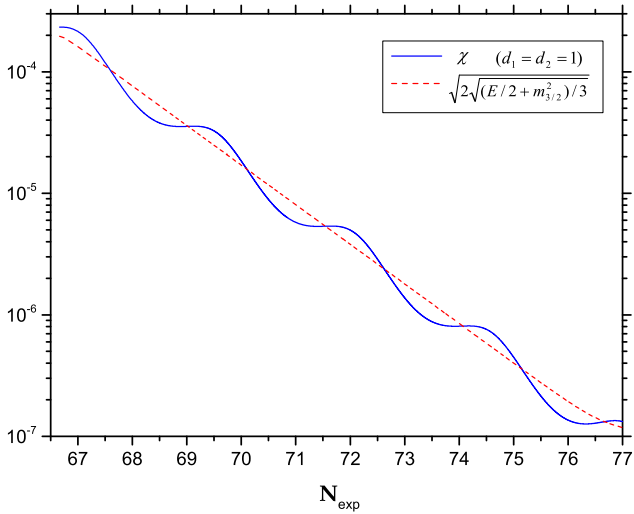,width=8.5cm}}
\caption{Postinflationary evolution of the field $\chi$ 
as a function of the number of e-foldings ${\bf N}_{\rm exp}$
for $m_{3/2}=10^{-14}$ and $d_1=d_2=1$. The moving position
in Eq.~(\ref{varmin}) of the minimum of the potential is also 
depicted.}
\label{fig1}
\end{figure}

\begin{figure}[t]
\centerline{\epsfig{file=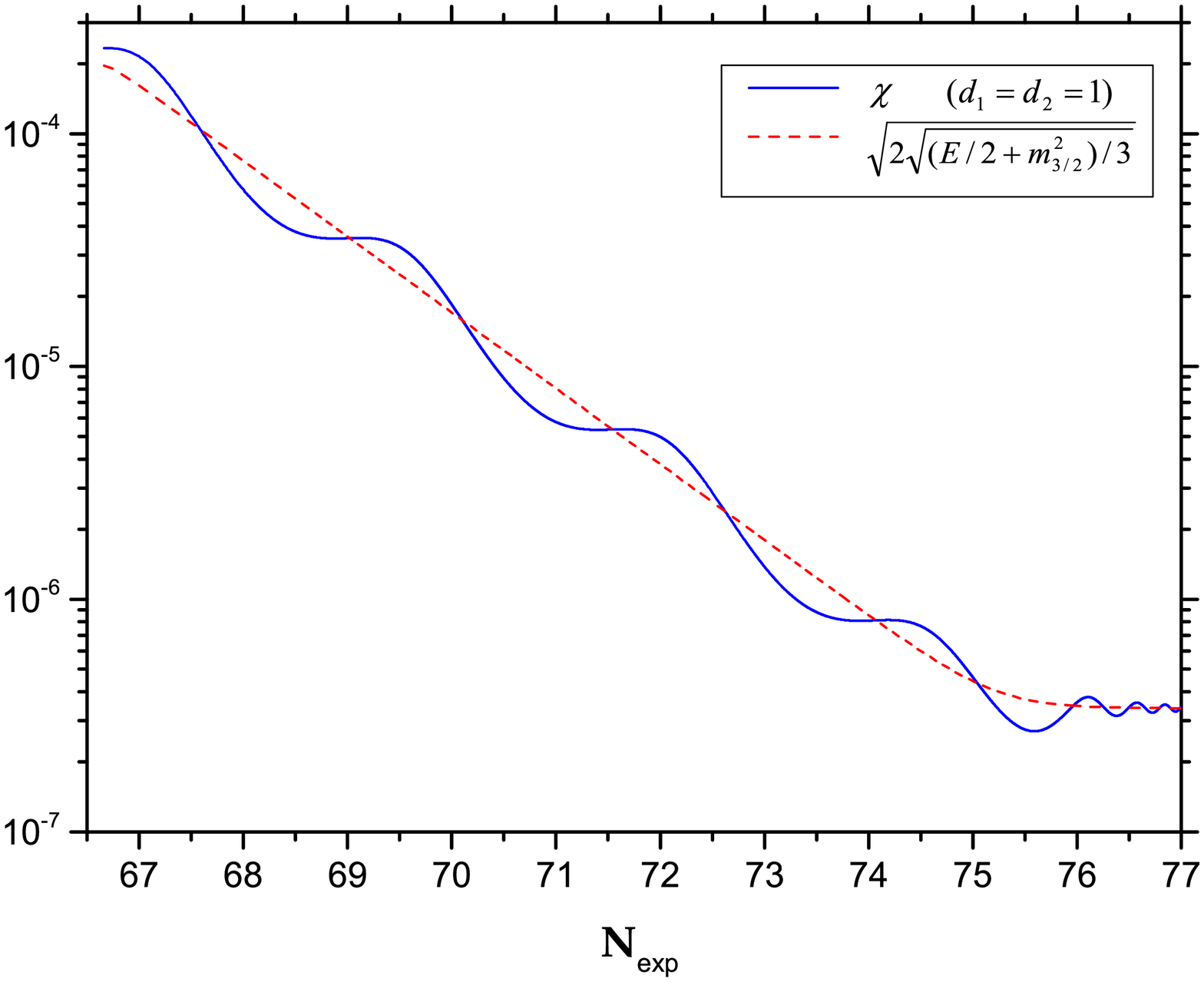,width=8.5cm}}
\caption{Postinflationary evolution of the field $\chi$ 
as a function of the number of e-foldings ${\bf N}_{\rm exp}$
for $m_{3/2}=10^{-13}$ and $d_1=d_2=1$. The moving position
in Eq.~(\ref{varmin}) of the minimum of the potential is also 
depicted.}
\label{fig2}
\end{figure}

\begin{figure}[t]
\centerline{\epsfig{file=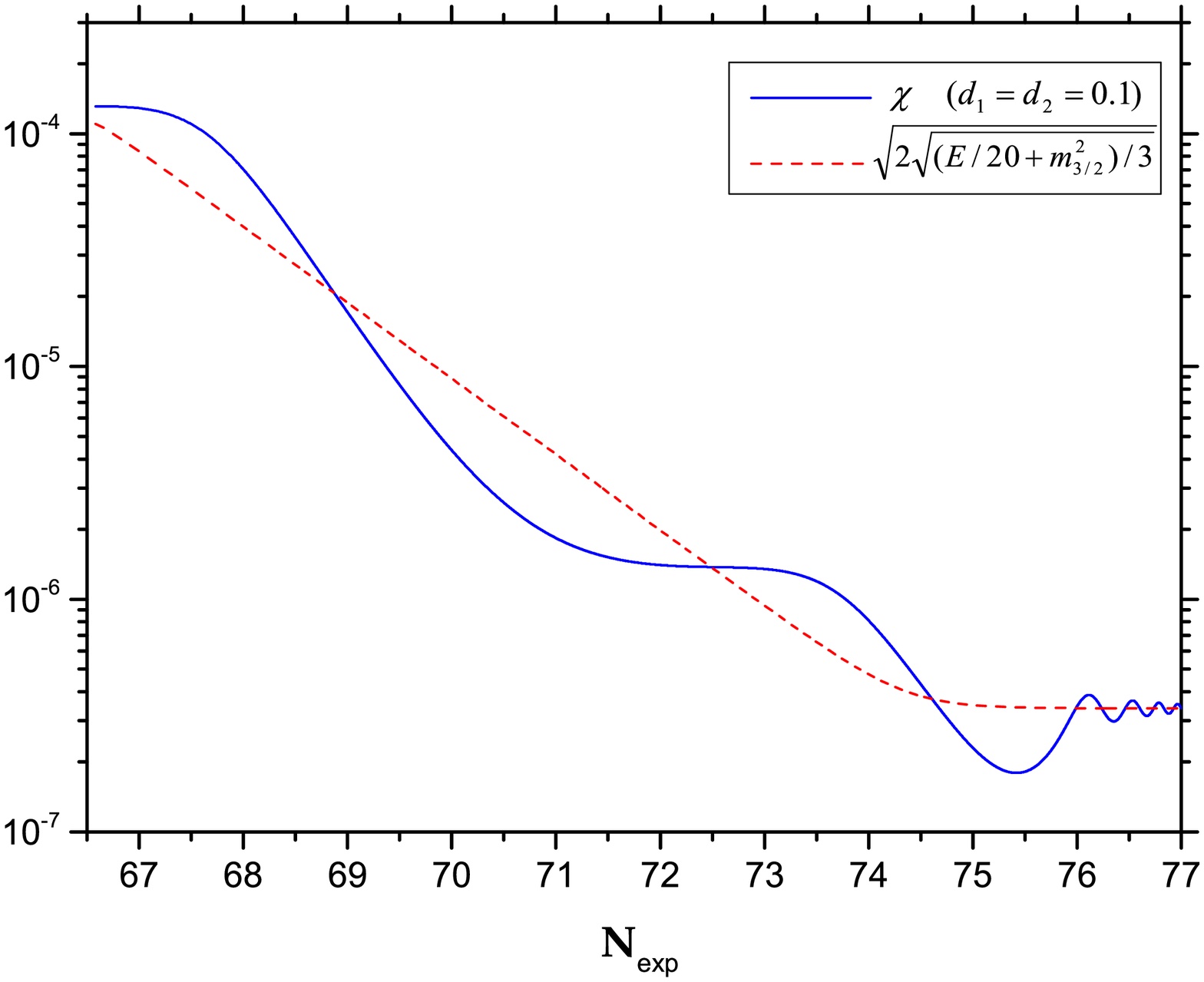,width=8.5cm}}
\caption{Postinflationary evolution of the field $\chi$
as a function of the number of e-foldings ${\bf N}_{\rm exp}$
for $m_{3/2}=10^{-13}$ and $d_1=d_2=0.1$. The moving position
in Eq.~(\ref{varmin}) of the minimum of the potential is also 
depicted.}
\label{fig4}
\end{figure}

\begin{figure}[t]
\centerline{\epsfig{file=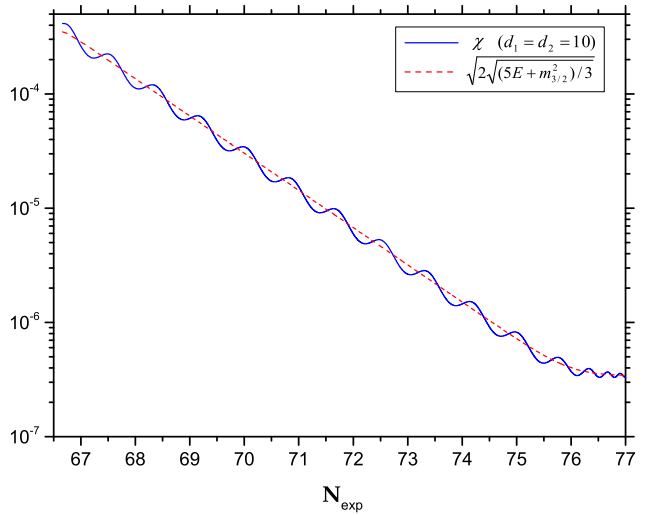,width=8.5cm}}
\caption{Postinflationary evolution of the field $\chi$
as a function of the number of e-foldings ${\bf N}_{\rm exp}$
for $m_{3/2}=10^{-13}$ and $d_1=d_2=10$. The moving position
in Eq.~(\ref{varmin}) of the minimum of the potential is also 
depicted.}
\label{fig5}
\end{figure}

So far we have not determined the coupling constant $\kappa$. 
Its value enters our considerations indirectly since it only 
controls the size of terms that we have neglected. From 
Eqs.~(\ref{deltah}) and (\ref{xend}), we see that inflation 
terminates via the waterfall mechanism provided $\kappa^2
\lesssim 8\pi^2M^2\simeq 3.72\times 10^{-4}$. In this case,
with $\kappa^2$ decreasing, $x_{\rm {end}}$ increases and, 
thus, the neglected term $(1/2\alpha)\ln (1-x_{\rm {end}})$ in 
Eq.~(\ref{N*}) increases in magnitude and, if included, leads 
to a larger reduction of ${\bf N}_*$. With increasing $\kappa^2$, 
instead, $\sigma_*^2$ increases and higher order terms in the 
potential which have not been taken into account start to play 
a role. As a suitable compromise, we may choose the value 
$\kappa=0.01$ yielding $|\sigma_*|=\sqrt{\delta_hx_*/\alpha}
\simeq 0.01378$, $x_{\rm {end}}\simeq 0.0248$, and $\epsilon_*
\simeq 4.22\times 10^{-9}$. So the tensor-to-scalar ratio
\begin{equation}
r=16\epsilon_*
\end{equation}
turns out to be about $6.76\times 10^{-8}$ implying that 
gravity waves are not observable in this case. Observable 
gravity waves in SUSY hybrid inflation have been discussed 
in Ref.~\cite{gravitywaves}. Taking into 
account the correction to ${\bf N}_*$ from  $x_{\rm {end}}$, we 
obtain ${\bf N}_*\simeq 50.1$. If we also include the effect of 
the term $\propto\sigma^4$ in the inflationary potential, 
${\bf N}_*$ is slightly reduced to ${\bf N}_*\simeq 49.9$, 
while $n_{\rm s}$ increases slightly to $n_{\rm s}
\simeq 0.9606$. 

To facilitate the theoretical analysis of the evolution of 
the field $\chi$ after the end of inflation, it is convenient 
to assume that $d_1=d_2=d$ (choosing e.g., $c_1=c_2=c_3$). 
In this case,
\begin{equation}
m_{\chi}^2 \simeq d\, V+m_{{3}/{2}}^2,
\end{equation}
provided that $V_{\chi} \ll V$. During the period following 
the end of inflation and before reheating, the energy density 
$E$ is dominated by the energy density of the rapidly 
oscillating massive fields $\sigma$ and $h$. Then, we may 
approximate $V$ in the above expression for $m_{\chi}^2$ with 
its average value
during one oscillation which is expected to be equal to $E/2$. 
Thus, we are led to the approximate expression
\begin{equation}
\label{varmin}
|\chi_{\rm m}|\simeq\sqrt{2}\left(\frac{d\, E/2+m_{{3}/{2}}^2}
{3\lambda^2}\right)^{\frac{1}{4}}
\end{equation}
for the position of the minimum of the potential 
$V_{\chi}$ in terms of the energy density 
$E$. As $d\, E/2$ approaches $m_{{3}/{2}}^2$, the decrease of 
$|\chi_{\rm m}|$ becomes slower until it reaches smoothly the 
value
\begin{equation}
\label{pqmin}
|\vev{\chi}|= \sqrt{2}|\vev{N}|\equiv f_{\rm a}= \sqrt{2}
\left(\frac{m_{{3}/{2}}^2}{3\lambda^2}\right)^{\frac{1}{4}},
\end{equation}
which determines the PQ breaking scale $f_{\rm a}$ (axion 
decay constant).

\begin{figure}[t]
\centerline{\epsfig{file=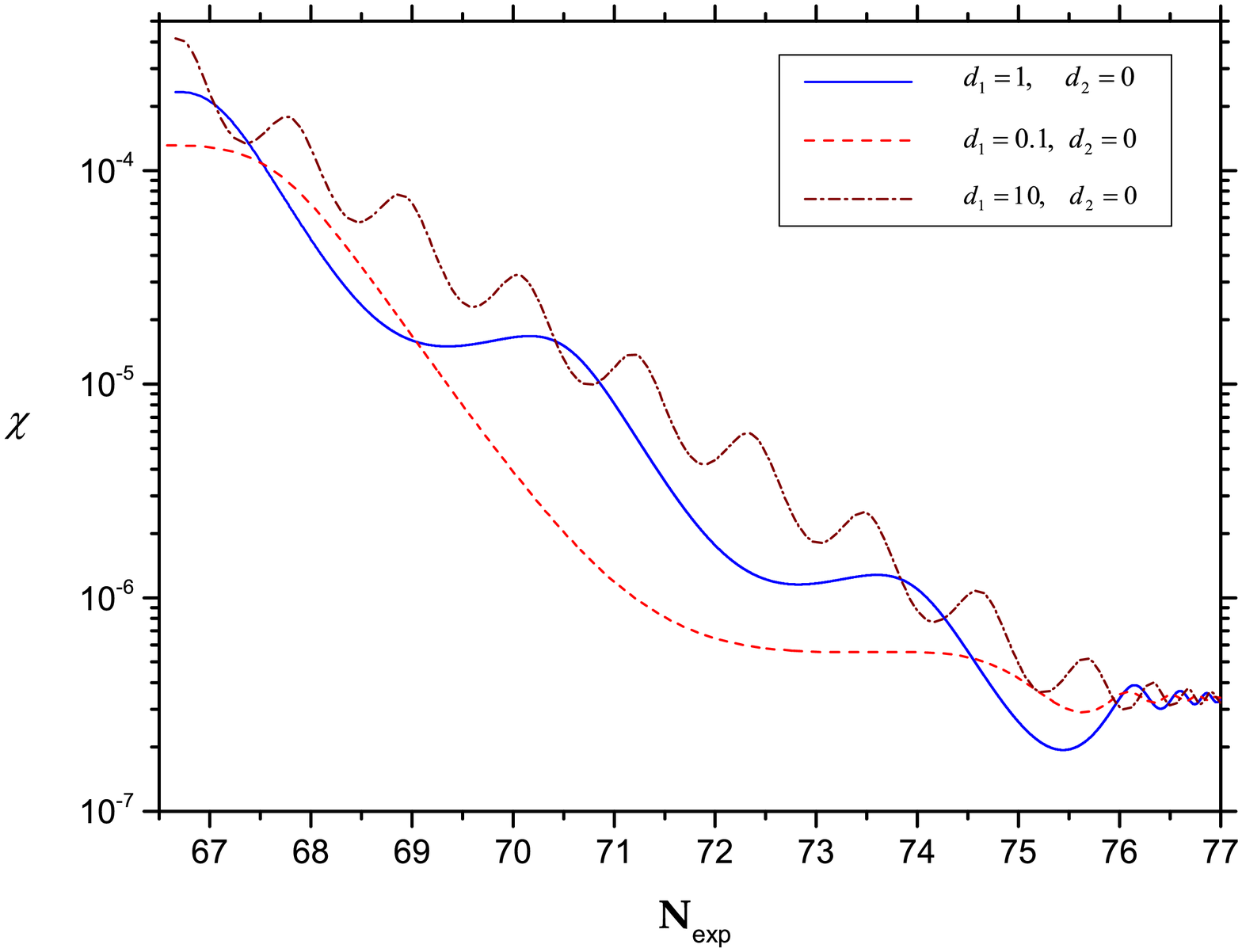,width=8.5cm}}
\caption{Postinflationary evolution of the field $\chi$
as a function of the number of e-foldings ${\bf N}_{\rm exp}$
for $m_{3/2}=10^{-13}$ and $d_1=1$, $0.1$, $10$ and $d_2=0$.}
\label{fig6}
\end{figure}

\begin{figure}[t]
\centerline{\epsfig{file=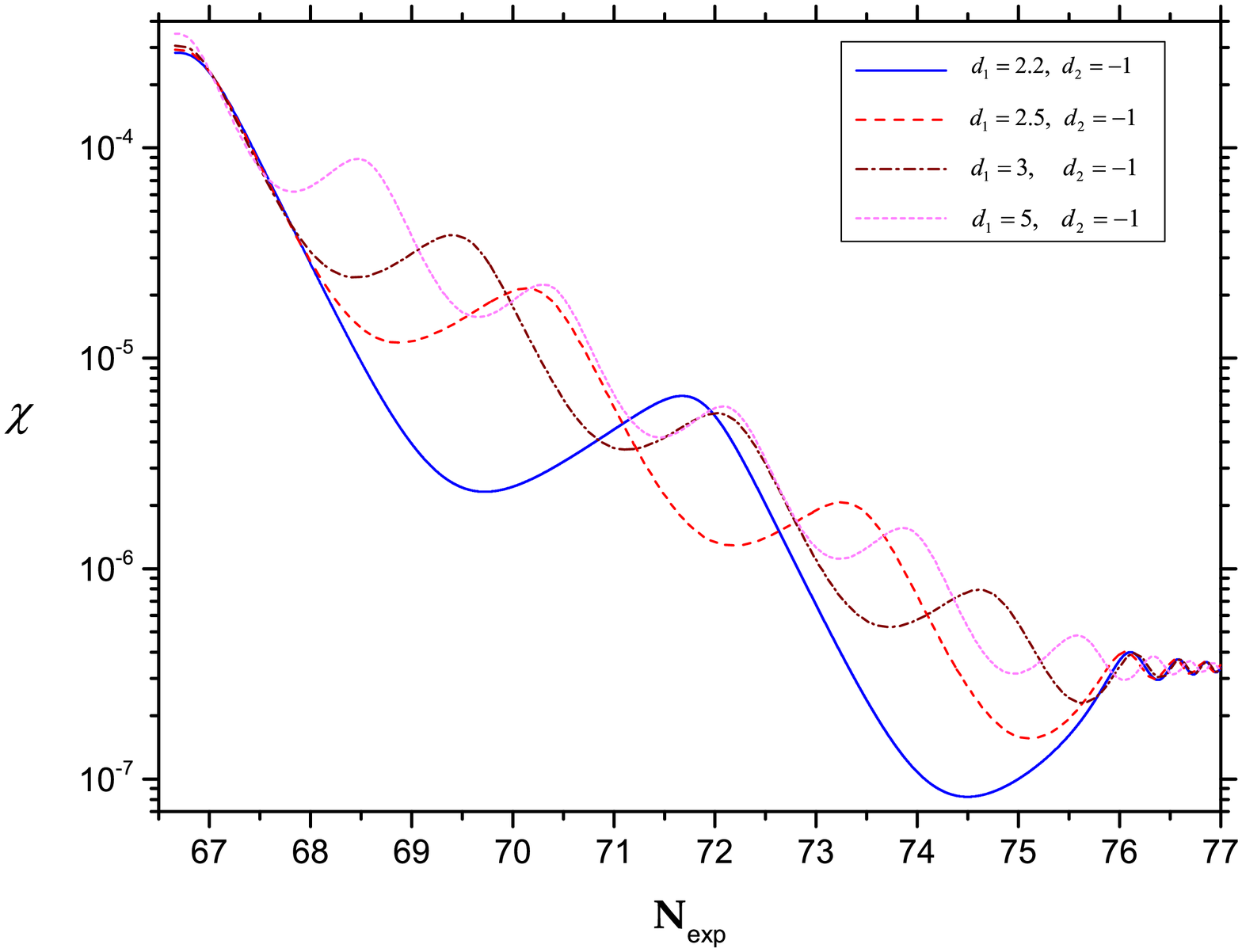,width=8.5cm}}
\caption{Postinflationary evolution of the field $\chi$
as a function of the number of e-foldings ${\bf N}_{\rm exp}$
for $m_{3/2}=10^{-13}$ and $d_1=2.2$, $2.5$, $3$, $5$ and 
$d_2=-1$.}
\label{fig7}
\end{figure}

In order to confirm our theoretical analysis concerning the 
postinflationary evolution of the field $\chi$, we solve
numerically the differential equations describing the 
dynamics of the system with potential energy density given 
in Eq.~(\ref{pot2}) and the following choice of parameters: 
$\kappa=0.01$, $\lambda=1$, $\alpha=1/150$, and $M=2.17 
\times 10^{-3}$. As initial conditions, we set: 
$\sigma=0.015$, $\dot\sigma=-1.86\times 10^{-12}$, 
$h=10^{-8}$, $\dot h=0$, $\chi=10^{-5}$, $\dot\chi=0$ 
(overdot denotes derivation with respect to cosmic time).
The initial velocity of $\sigma$ is the actual value of 
its velocity on the inflationary path, which was determined 
numerically. We followed the evolution of the system for 77 
e-foldings of expansion out of which approximately 66.7 
correspond to inflationary expansion. Notice that the 
initial value of $\sigma$ is larger than $|\sigma_*|$. The 
value of the final energy density after 77 e-foldings of 
expansion is close to $10^{-28}$. 

In Figs.~\ref{fig1}-\ref{fig5}, we plot the values of the 
field $\chi$ and the varying value of $|\chi_{\rm m}|$ in 
Eq.~(\ref{varmin}) as functions of the number of 
e-foldings ${\bf N}_{\rm exp}$ only for the period following 
the end of inflation. Note that the number of e-foldings 
${\bf N}_{\rm exp}$ is measured starting from the cosmic time 
at which the initial conditions were imposed. In 
Fig.~\ref{fig1}, $m_{3/2}=10^{-14}$ while in all other figures 
$m_{3/2}=10^{-13}$. Moreover, in Figs.~\ref{fig1} 
and \ref{fig2}, $d_1=d_2=1$, in Fig.~\ref{fig4}, 
$d_1=d_2=0.1$, and, in Fig.~\ref{fig5}, $d_1=
d_2=10$. We see that $\chi$ oscillates around the moving 
position $\chi_{\rm m}$ of the minimum in Eq.~(\ref{varmin}) 
adjusting faster to it when $d$ is larger. 
Also, in Fig.~\ref{fig1}, $\chi_{\rm m}$ after 77 e-foldings 
has just started approaching the value of the PQ vacuum of 
Eq.~(\ref{pqmin}), while in all other figures it essentially
has reached the value of Eq.~(\ref{pqmin}) and the amplitude 
of the oscillations of $\chi$ has been severely reduced. Our 
numerical findings are in accord with our theoretical 
expectations. It should be noted that, reducing the parameter 
$\lambda$ below 1, $\chi_{\rm m}$ is just rescaled by a factor 
$\lambda^{-1/2}$ as one can see from Eq.~(\ref{potmin}). 
Consequently, in all these figures, $\chi$ is practically 
just multiplied by the same factor. This is confirmed 
numerically too.

Encouraged by the good agreement of the numerical results 
with our theoretical expectations so far, we also consider 
the limiting case $d_2=0$, but insisting that $d_1>0$ in 
order that $|\chi|$ have an appreciable value during 
inflation. In Fig.~\ref{fig6}, we plot the postinflationary 
evolution of the field $\chi$ versus ${\bf N}_{\rm exp}$ for 
$m_{3/2}=10^{-13}$ and three values of $d_1$, namely 
$d_1=1$, $0.1$, $10$ and a common $d_2=0$. We see that $\chi$ 
decreases oscillating with larger frequency for larger 
values of $d_1$ (larger $m_{\chi}^2$) until it approaches 
the PQ vacuum where the amplitude of the oscillations dies 
out. We finally allow in the numerical investigation 
even negative values of $d_2$, but always with $d_1>0$. 
Such values of $d_2$ can be obtained by assuming that 
$c_2$, $c_3<1$. In Fig.~\ref{fig7}, we made the choices 
$d_1=2.2$, $2.5$, $3$, $5$ with a common $d_2=-1$ 
(corresponding to $c_2=c_3=0$) and the same value of 
$m_{3/2}$ as in Fig.~\ref{fig6}. The emerging picture is 
roughly the same. Notice 
that in this case there are also negative contributions to 
$m_{\chi}^2$, which reduce its value and with it the absolute 
value of the position of the variable minimum of $V_{\chi}$ 
during the postinflationary era. As a consequence, the 
oscillating field $\chi$ acquires both positive and negative 
values during its oscillation unless $d_1$ is sufficiently 
large. We found that this is avoided for $d_1$ larger than 
$\simeq 2.2$.

\section{Relic density of the PQ fields}
\label{relic}

We will now estimate the relic densities and decay times of 
the particles contained in the PQ system. The relevant 
superpotential term is
\begin{equation}
\delta W_1=\lambda N^3\bar{N}.
\label{deltaW1}
\end{equation}
The resulting potential in global SUSY is 
\begin{equation}
V=\lambda^2 |N|^4(|N|^2+9|\bar{N}|^2)-m_{N}^2|N|^2+m_{\bar{N}}^2
|\bar{N}|^2,
\label{pot}
\end{equation}
where $-m_{N}^2=\varepsilon_1m_{3/2}^2$, $m_{\bar{N}}^2=
\varepsilon_2m_{3/2}^2$
are the soft masses squared of the fields $N$, $\bar{N}$ and, for 
simplicity, the soft A term is set to zero. In order to find the 
extrema of this potential we take its derivatives with respect to 
$|N|$, $|\bar{N}|$ and put them equal to zero:
\begin{eqnarray}
\frac{\partial V}{\partial |N|}&=&2(3\lambda^2 |N|^4-m_{N}^2)|N|=0,
\nonumber\\
\frac{\partial V}{\partial |\bar{N}|}&=&2(9\lambda^2|N|^4+m_{\bar{N}}^2)
|\bar{N}|=0.
\end{eqnarray}
We find two extrema: one at $|N|=|\bar{N}|=0$, which is a local 
maximum, and one at
\begin{equation}
|N|=\left(\frac{m_N^2}{3\lambda^2}\right)^{\frac{1}{4}}\equiv\vev{N}, 
\quad \quad |\bar{N}|=0,
\label{vacuum}
\end{equation} 
which is the absolute minimum with potential energy density
\begin{equation}
V_0=-\frac{2}{3}m_N^2\left(\frac{m_N^2}{3\lambda^2}
\right)^{\frac{1}{2}},
\end{equation}
and corresponds to the PQ vacuum in Eq.~(\ref{pqmin}).

Replacing $N=\vev{N}+\delta N$ in the potential in Eq.~(\ref{pot}) and 
keeping only terms quadratic in $\delta N$ and $\bar{N}$, we find
\begin{eqnarray}
V&=&(-m_N^2+3\lambda^2\vev{N}^4)|\delta{N}|^2 +
3\lambda^2\vev{N}^4(\delta N+\delta N^*)^2\nonumber\\& &
+(3m_N^2+m_{\bar{N}}^2)|\bar{N}|^2.
\end{eqnarray}  
Writing $\delta N=(\delta\chi+ia)/\sqrt{2}$, we obtain a real scalar 
field $\delta\chi$ with mass squared $m_{\delta\chi}^2=4m_N^2$ which 
is the saxion, a 
massless real scalar field $a$ which is the axion, and an extra  
complex scalar field $\bar{N}=(\bar{\chi}+i\bar{a})\sqrt{2}$ with 
mass squared $m_{\bar{\chi}}^2=3m_N^2+m_{\bar{N}}^2$.  

The saxion predominantly decays into a pair of MSSM Higgsinos 
$\tilde{h}_u$, $\tilde{h}_d$ (tilde denotes SUSY partner) 
via the Yukawa coupling 
\begin{equation} 
-2\sqrt{2}\,\frac{\mu}{f_{\rm a}}\,\delta N\tilde{h}_u\tilde{d}_d+h.c.,
\label{saxionhiggsino}   
\end{equation}
which originates from the superpotential term
\begin{equation}
\delta W_2=\frac{1}{2}\lambda_\mu N^2 h^2,
\label{deltaW2}
\end{equation}
where $h=(h_u, h_d)$ with $h_u$, $h_d$ being the electroweak Higgs 
superfields which couple, respectively, to the up-type and 
down-type quarks. It is important to note that 
the superpotential term in Eq.~(\ref{deltaW2}) is \cite{nilles} 
also responsible for the generation of the MSSM $\mu$ term, 
with $\mu=\lambda_\mu\vev{N}^2$ after the spontaneous 
breaking of the PQ symmetry. The origin and the magnitude 
of the MSSM $\mu$ term is nicely attributed to the PQ 
mechanism. 

The saxion decay width is estimated to be
\begin{equation}
\Gamma_{\rm d}\simeq \frac{1}{\pi}\left(\frac{\mu}{f_{\rm a}}
\right)^2m_{\delta\chi}.
\label{Gamma} 
\end{equation}  
The energy density $\rho_{\delta\chi}$ from the coherent saxion 
field oscillations about the vacuum behaves like pressureless 
matter. At cosmic time $t_{\rm d}=\Gamma^{-1}_{\rm d}$ of saxion 
decay this density is 
\begin{equation}
\rho_{\delta\chi}^{\rm d}=\rho_{\delta\chi}^{\rm r}\left(\frac{t_{\rm r}}
{t_{\rm d}}\right)^{\frac{3}{2}},
\label{rhod}
\end{equation}  
where $\rho_{\delta\chi}^{\rm r}$ is the energy density of saxion 
oscillations at the time $t_{\rm r}$ of reheating following 
inflation. Here, we assumed that the universe is radiation 
dominated after reheating and at least up to the time of 
saxion decay. This will be justified {\it a posteriori}. 
Note also that, as it turns out, $t_{\rm r}<t_{\rm d}$ (see 
below).   

Our numerical calculations show that the amplitude of coherent 
saxion oscillations about the PQ vacuum at the onset of these 
oscillations at cosmic time $t_{\rm osc}\sim m_{\delta\chi}^{-1}$ 
is of order $f_{\rm a}$. Consequently, the corresponding initial 
energy density is 
\begin{equation}
\rho_{\delta\chi}^{\rm osc}\sim \frac{1}{2}\,m_{\delta\chi}^{2}
f_{\rm a}^2,
\label{rhoosc}
\end{equation}
which implies that 
\begin{equation}
\rho_{\delta\chi}^{\rm r}\sim \frac{1}{2}\,m_{\delta\chi}^{2}
f_{\rm a}^2\left(
\frac{t_{\rm osc}}{t_{\rm r}}\right)^2,
\label{rhor}
\end{equation}
since the universe is matter dominated between $t_{\rm osc}$ 
and $t_{\rm r}$. Note that $t_{\rm osc}<t_{\rm r}$ (see below). 

From Eqs.~(\ref{rhod}) and (\ref{rhor}), we then find that
\begin{equation}
\rho_{\delta\chi}^{\rm d}\sim \frac{1}{2\pi^2}\,
m_{\delta\chi}^{2}f_{\rm a}^2
\left(\frac{t_{\rm d}}{t_{\rm r}}\right)^\frac{1}{2}\left(
\frac{\mu}{f_{\rm a}}\right)^4.
\label{rho}
\end{equation}
For $f_{\rm a}=10^{11}~{\rm GeV}$ and $\mu=m_{\delta\chi}=1~{\rm TeV}$,
Eq.~(\ref{Gamma}) yields $t_{\rm d}\simeq 3.14\times 10^{13}~
{\rm GeV}^{-1}$. The cosmic time at reheating is given by 
\begin{equation}
t_{\rm r}=\frac{3\,\sqrt{5}}{\sqrt{2}\,\pi \,T_{\rm r}^{2}\,
g_*(T_{\rm r})^{\frac{1}{2}}},
\end{equation}
where $T_{\rm r}$ is the reheat temperature and $g_*(T_{\rm r})$ 
denotes the effective number of massless degrees of 
freedom at $T_{\rm r}$. For $T_{\rm r}=10^9~{\rm GeV}$ and 
$g_*(T_{\rm r})=228.75$,
this equation gives $t_{\rm r}\simeq 0.24~{\rm GeV}^{-1}$. From 
Eq.~(\ref{rho}) we then obtain $\rho_{\delta\chi}^{\rm d}\simeq 58~
{\rm GeV}^4$, which compared with the radiation energy density 
\begin{equation}
\rho_{\rm rad}^{\rm d}=\frac{\pi^2}{30}\,g_*(T_{\rm d})\,
T_{\rm d}^4\simeq 4.5\times 10^9~{\rm GeV}^4
\label{rad}
\end{equation}
at the temperature $T_{\rm d}$ of saxion decay, implies that
\begin{equation}
\rho_{\delta\chi}^{\rm d}\ll \rho_{\rm rad}^{\rm d}.
\end{equation}
Thus, the energy density of coherent saxion oscillations never 
dominates the energy density of the universe. In Eq.~(\ref{rad}), 
$g_*(T_{\rm d})$ is the effective number of massless degrees of 
freedom at temperature $T_{\rm d}$, which is given by
\begin{equation}
T_{\rm d}=\frac{3^{\frac{1}{2}}5^{\frac{1}{4}}}
{2^{\frac{1}{4}}\pi^{\frac{1}{2}}t_{\rm d}^{\frac{1}{2}}
g_*(T_{\rm d})^{\frac{1}{4}}}.
\end{equation}
In our numerical example, this formula gives $T_{\rm d}\simeq 
110~{\rm GeV}$ for $g_*(T_{\rm d})=95.25$. The freeze-out 
temperature of the LSP, which is assumed to be the lightest 
neutralino, is $T_{\rm f}\simeq m_{\rm LSP}/25$ 
\cite{freeze} (for a review see Ref.~\cite{neutralino}), 
where  $m_{\rm LSP}$ is the mass of the LSP. The saxions 
decay before the LSPs freeze out provided that
\begin{equation}
m_{\rm LSP}\lesssim 1.65~{\rm TeV},
\end{equation} 
and so they do not affect the usual neutralino relic 
abundance since they remain subdominant until their decay. 

The above argument is not complete though, since so far we 
only considered the energy density of the coherently oscillating
saxion field and ignored the thermally produced saxions. In the 
present case, the thermal saxion production is dominated by 
processes involving the saxion-Higgsino-Higgsino coupling in
Eq.~(\ref{saxionhiggsino}). Unfortunately, to the best of our 
knowledge, there is no exact calculation of the relic thermal 
saxion abundance $Y_{\delta\chi}\equiv n_{\delta\chi}/s$ in this 
case ($n_{\delta\chi}$ is the thermal saxion number density and 
$s$ the entropy density). However, it is expected 
\cite{saxiondensity} to be similar to the thermal axino 
$\tilde{a}$ abundance \cite{axinodensity} 
\begin{equation}
Y_{\tilde{a}}\simeq 10^{-5}\left(\frac{f_{\rm a}}
{10^{11}~{\rm GeV}}\right)^{-2}\left(\frac{\mu}{10^{3}~
{\rm GeV}}\right)^2,
\end{equation}    
which yields $Y_{\tilde{a}}\simeq 10^{-5}$ for our numerical 
inputs. Consequently, the thermal saxion energy density 
$\rho_{\delta\chi}^{\rm d}$ at saxion decay is estimated to 
be given by 
\begin{equation}
\frac{\rho_{\delta\chi}^{\rm d}}{\rho_{\rm rad}^{\rm d}}\simeq 
\frac{4}{3}\frac{Y_{\delta\chi} m_{\delta\chi}}{T_{\rm d}}\sim 
10^{-4} 
\ll 1
\end{equation}
and the thermally produced saxions also remain subdominant
until saxion decay. Note that, in this model, the thermal 
production of saxions or axinos mostly takes place at low
cosmic temperatures of order $\mu$ and is independent from the 
reheat temperature.  

The scalar sector of our model contains not only the saxion 
and axion fields, but also the $\bar{N}$ complex Higgs field. 
Our numerical simulations show that this field,
for $A=0$, approaches zero very quickly with its coherent 
oscillations about the vacuum being practically of vanishing 
amplitude. Consequently, these oscillations do not contribute
to the energy density of the universe. On the contrary, the 
thermally produced $\bar{N}$ scalar particles can contribute 
to the energy density. In the present model, the thermal 
production of $\bar{N}$'s is dominated by processes involving 
the coupling 
\begin{equation}
2\sqrt{2}\,\frac{\mu}{f_{\rm a}}\,\frac{m_{\tilde{a}}}
{m_{\bar{\chi}}}\,m_{\bar{\chi}} \bar{N} h_u h_d + h.c.,
\label{extra}
\end{equation}      
where $m_{\tilde{a}}=\sqrt{3}m_N$ is the axino mass (see 
below). This coupling originates from the cross F term
between $\delta W_1$ in Eq.~(\ref{deltaW1}) and $\delta W_2$ 
in Eq.~(\ref{deltaW2}) and, for 
$m_{\tilde{a}}\sim m_{\bar\chi}$, has practically the same 
dimensionless coupling constant as the Yukawa coupling in 
Eq.~(\ref{saxionhiggsino}). Moreover, the $\bar{N}$ Higgs 
field predominantly decays into a pair of ordinary Higgs 
fields via the coupling in Eq.~(\ref{extra}). As a 
consequence, both the decay temperature and thermal 
abundance of $\bar{N}$ are similar to those of the saxion 
and this field also decays well before the freeze out of 
neutralinos without dominating the energy density of the 
universe.

Next let us turn to the discussion of the fermionic sector 
of the PQ system. The superpotential term $\delta W_1$
gives rise to the following Yukawa coupling
\begin{equation}
-3\lambda \vev{N}^2\tilde{N}\tilde{\bar{N}}+h.c.\,.
\end{equation}
Substituting $\vev{N}$ from Eq.~(\ref{vacuum}), 
we obtain a four component massive axino field $\tilde{a}$ with 
Dirac mass $m_{\tilde{a}}=\sqrt{3}m_N$. The dominant vertices
involving this field come from the superpotential term 
$\delta W_2$ and are given by
\begin{equation}
-2\sqrt{2}\,\frac{\mu}{f_{\rm a}}\,\tilde{N}\tilde{h}_uh_d
-2\sqrt{2}\,\frac{\mu}{f_{\rm a}}\,\tilde{N}\tilde{h}_dh_u+h.c.\,.
\end{equation}
The axino predominantly decays into an ordinary Higgs field
and a Higgsino with a decay width similar to that of the 
saxion, again well before the neutralino freeze out and remains
subdominant. 

The overall conclusion is that the saxion, $\bar{N}$ Higgs 
field, and axino do not affect the universe in any essential 
way. In particular they do not alter the LSP relic abundance. 

Finally, let us note that there will be also thermal axion 
production \cite{thermalaxions}. In our case, this is again 
dominated by processes involving the coupling in 
Eq.~(\ref{saxionhiggsino}). Consequently, the resulting 
abundance is expected to be similar to the one of thermal 
saxions. The contribution of the thermal axions to the 
effective number of neutrinos turns out to be tiny. One 
reason leading to this result is the quite large number of 
degrees of freedom at high temperatures due to supersymmetry. 

\section{Axion Isocurvature Perturbations}
\label{sec:iso}

The axion isocurvature perturbation can be calculated as 
follows \cite{lyth}. The axion field during inflation is massless 
and thus acquires a perturbation which, at horizon exit of the 
pivot scale $k_*=0.05~{\rm Mpc}^{-1}$, is given by 
\begin{equation}
\delta\hat{a}=\frac{H_*}{2\pi},
\end{equation}
where $H_*$  is the corresponding Hubble parameter. The value of 
the axion field at horizon crossing of $k_*$ is $\hat{a}=
\theta\chi_*$, where $\theta$ is the so-called initial misalignment 
angle, i.e. the phase of the complex axion field $N$ during 
inflation, and $\chi_*$ is the corresponding value of the real scalar   
field $\chi$. Consequently, the perturbation in the misalignment angle is 
\begin{equation}
\delta\theta=\frac{H_*}{2\pi\chi_*}.   
\end{equation}   

At the QCD phase transition, the axion acquires a mass $m_{a}$ and
starts performing coherent oscillations with an initial amplitude
$a=\theta f_{\rm a}$ and energy density 
\begin{equation}
\rho_{a}=\frac{1}{2}m_{a}^2 a^2.
\end{equation}
The perturbation in $\theta$ then translates into a perturbation
in the initial amplitude
\begin{equation}
\delta a= \delta\theta f_{\rm a}=\frac{H_*f_{\rm a}}{2\pi\chi_*},
\end{equation}
yielding a perturbation $\delta n_a$ in the axion number density 
$n_a$. The resulting axion isocurvature perturbation is 
\begin{equation}
{\cal S}_a=\frac{\delta n_a}{n_a}=\frac{\delta \rho_a}{\rho_{a}}=
2\frac{\delta a}{a}=\frac{H_*}{\pi \theta\chi_*},
\end{equation}
and is completely uncorrelated with the curvature perturbation. 
The overall isocurvature perturbation generated by the axions 
is then  
\begin{equation}
{\cal S}=\frac{H_*}{\pi \theta\chi_*}R_a,
\label{iso}
\end{equation}
where $R_a$ is the axion fraction of cold dark matter in the 
universe.

The recent Planck satellite data indicate \cite{planck} that
the primordial isocurvature fraction
\begin{equation}
\beta_{\rm iso}(k_*)\equiv \frac{{\cal P_{II}}(k_*)}
{{\cal P_{RR}}(k_*)+{\cal P_{II}}(k_*)}
\end{equation} 
at the pivot scale $k_*$, in the case of uncorrelated 
isocurvature perturbations with spectral index equal to 
unity (i.e. with ${\cal P_{II}}$ being scale independent), 
should be less than about 0.038 at $95\%$ confidence 
level. Here 
${\cal P_{RR}}$ and  ${\cal P_{II}}$ are the adiabatic 
and isocurvature power spectra respectively. This 
result then implies that, at $95\%$ confidence level, 
\begin{equation}
{\cal S}\lesssim 0.1987\,A_{\rm s}(k_*)^\frac{1}{2}\simeq
9.35\times 10^{-6},
\label{bound}
\end{equation}
where $A_{\rm s}(k_*)\simeq 2.215 \times 10^{-9}$ is the 
amplitude of the scalar power spectrum.

The axion fraction of cold dark matter is given by 
$R_{a}=\Omega_{a} h^2/\Omega_{\rm CDM} h^2$, where the relic 
axion abundance 
is \cite{gondolo}
\begin{equation}
\Omega_a h^2\simeq 0.236\left(\frac{f_{\rm a}/{\cal N}}
{10^{12}~{\rm GeV}}\right)^{\frac{7}{6}}\langle{{\cal N}^2
\theta^2f({\cal N}\theta)}\rangle,
\label{omega}
\end{equation}
and the relic abundance of cold dark matter $\Omega_{\rm CDM}h^2
\simeq 0.12$ \cite{planck}. The misalignment angle $\theta$ 
lies \cite{curvaton} in the interval $[-\pi/{{\cal N}}, +
\pi/{{\cal N}}]$, where ${\cal N}$ is the absolute value of the 
sum of the PQ charges of all fermionic color (anti)triplets and 
all $\theta$'s in this interval are equally probable. In the 
simplest models such as ours, ${\cal N}=6$. The function 
$f({\cal N}\theta)$ accounts for the anharmonicity of the axion 
potential and the average 
$\langle{{\cal N}^2\theta^2f({\cal N}\theta)}\rangle$ is 
evaluated in the above interval. 

Note that this same ${\cal N}$ determines the $Z_{\cal N}$ 
subgroup of $U(1)_{\rm PQ}$ which remains explicitly unbroken by 
instantons. This discrete subgroup, however, is spontaneously 
broken by the PQ symmetry breaking VEV $\vev{N}$ of $N$. This 
would lead \cite{sikivie} to a cosmologically disastrous domain 
wall production if the spontaneous breaking of the PQ symmetry 
occurred after inflation. Avoidance of this catastrophe would 
then require extending \cite{axionwalls} the simplest model so 
that such spontaneously broken discrete symmetries do not appear. 
Fortunately, in our case here, the PQ symmetry is already 
spontaneously broken during inflation and remains so thereafter.  
Thus, no domain wall problem arises and there is no need of 
extending the simplest model.    

The average $\langle{{\cal N}^2\theta^2f({\cal N}\theta)}
\rangle$ evaluated in the interval $[-\pi/{\cal N}, 
+\pi/{\cal N}]$ turns out to be \cite{gondolo} about $8.77$. 
Consequently, for $f_{\rm a}=10^{11}~{\rm GeV}$, we 
obtain $R_a\simeq 0.14$. For $H_*\simeq H_{\rm inf}=\kappa 
M^2/\sqrt{3}\simeq 2.72\times 10^{-8}$ in our numerical example 
(recall that $\kappa=0.01$ and $M\simeq 2.17\times 10^{-3}$), 
Eq.~(\ref{iso}) gives 
\begin{equation}
{\cal S}\simeq2.89\times 10^{-8}\,\frac{R_a}{\chi_*}\simeq 
4\times 10^{-9}\,\frac{1}{\chi_*}
\label{Schi*}
\end{equation} 
($H_{\rm inf}$ is approximately the Hubble parameter during 
inflation and $\theta$ is taken to be about $0.3$, which is 
the square root of the mean value of $\theta^2$ in the above 
interval). The bound in Eq.~(\ref{bound}) then implies 
\begin{equation}
\chi_*\gtrsim 4.28\times 10^{-4}.
\label{chibound}
\end{equation} 
However, as we saw in Sec.~\ref{trans}, for $d_1=1$, the value 
of $\chi$ during inflation is
\begin{equation}
\chi_*\simeq |\chi_{\rm inf}|\simeq\sqrt{2}\left( \frac{V_*}
{3\lambda^2}\right)^{\frac{1}{4}}\simeq 
\frac{2.33\times 10^{-4}}{\lambda^\frac{1}{2}},
\label{chi*inf}
\end{equation}
where $V_*\simeq\kappa^2M^4$ is the potential energy density 
during inflation. So the bound in Eq.~(\ref{chibound}) is 
satisfied provided that $\lambda\lesssim 0.3$, which is 
quite natural and, using Eq.~(\ref{vacuum}), gives $m_N
\lesssim 1069~{\rm GeV}$.

Allowing smaller values of the parameter $\lambda$ and, thus, 
bigger values of the axion decay constant $f_{\rm a}$, we 
can achieve much bigger values of the axion fraction of cold 
dark matter in the universe. For example, for $f_{\rm a}=4
\times 10^{11}~{\rm GeV}$, Eq.~(\ref{omega}) gives $R_a\simeq 
0.73$. Using Eq.~(\ref{Schi*}), we then find that the bound 
in Eq.~(\ref{bound}) is satisfied for $\chi_*\gtrsim 2.26
\times 10^{-3}$. From Eqs.~(\ref{vacuum}) and (\ref{chi*inf}), 
we see that this requires $\lambda\lesssim 0.01$ and 
$m_N\lesssim 570~{\rm GeV}$ or, equivalently, saxion mass 
$m_{\delta\chi}\lesssim 1140~{\rm GeV}$. So the axion fraction 
of dark matter can be very sizable with natural values of the 
parameters. In such a case, axions may be detectable in future 
microwave cavity experiments \cite{cavity}.

\section{Summary}
\label{concl}

We have provided a simple realistic SUSY hybrid inflation model 
which incorporates the PQ mechanism for solving the strong CP 
problem and intimately links the resolution of the MSSM $\mu$ 
problem with axion physics. We investigated how the PQ transition 
proceeds in this model by closely monitoring the scalar fields 
that accompany the axion, such as the saxion, during and after 
inflation to insure that no cosmological problems are encountered. 

A potential problem is the generation of unacceptably large axion 
isocurvature perturbations. We showed that this problem can be 
avoided provided that the value of the PQ field during inflation 
is suitably large, which can be achieved by using appropriate higher 
order terms in the K\"{a}hler potential. These terms generate a 
suitable negative mass-squared term for the PQ field and shift it 
to appropriately large values during inflation. The value of the 
PQ field also remains large during the subsequent inflaton 
oscillations, but gradually drifts to the desired lower energy 
PQ vacuum. We find that the scalar spectral index $n_{\rm s}$ can 
easily satisfy the observational bounds, while the tensor-to-scalar 
ratio $r$ turns out to be negligible. So no measurable gravity waves 
are predicted in this simple model. The PQ symmetry is already 
spontaneously broken during inflation and remains so thereafter. 
As a consequence, the axion domain walls are inflated away and thus 
the potential cosmological catastrophe is avoided. 

We estimated the relic saxion energy density in the universe due 
to saxion oscillations about the PQ vacuum as well as to the thermal 
production of saxions. We find that the saxions remain subdominant 
until their final decay. The same is true for the complex $\bar{N}$ 
Higgs field which is contained in the PQ system of the model. The 
axino, which is a heavy four component Dirac fermion in this model, 
also remains subdominant and decays well before the freeze out of 
the LSPs. The neutralino relic density is not affected and, 
depending on the axion decay constant $f_{\rm a}$ and the magnitude 
of the $\mu$ parameter, the axions and/or the lightest SUSY particle 
compose the dark matter in the universe.

Before concluding the following remarks are in order. Firstly,  
we constructed our model based on the left-right symmetric gauge 
group $SU(3)_{\rm c}\times SU(2)_{\rm L}\times SU(2)_{\rm R}\times 
U(1)_{B-L}$ whose spontaneous breaking does not produce 
\cite{trotta} magnetic monopoles or cosmic strings. Thus, with 
this choice standard SUSY 
hybrid inflation works fine with the spontaneous breaking of the 
gauge symmetry occurring at the end of inflation. With a different 
choice of gauge group such as the Pati-Salam group $SU(4)_{\rm c}
\times SU(2)_{\rm L}\times SU(2)_{\rm R}$ whose spontaneous 
breaking leads \cite{magg} to the production of magnetic 
monopoles, our axion model is 
successfully implemented by utilizing the shifted \cite{shift} or 
smooth \cite{smooth} variant of hybrid inflation which inflates 
away the magnetic monopoles. Secondly, in this class of models 
the reheat process after inflation yields right handed neutrinos 
and sneutrinos whose subsequent decay yields the observed baryon 
asymmetry of the universe via thermal \cite{thermallepto} or 
non-thermal \cite{nonthermalepto} leptogenesis. In contrast, 
if the MSSM $\mu$ problem is resolved via the superpotential 
coupling $S h^2$, as suggested in Ref.~\cite{dvali1}, 
the reheat temperature from this coupling lies well above 
the $10^7-10^9~{\rm GeV}$ bound from considerations of the 
gravitino \cite{gravitino} and one is led \cite{split} to split 
SUSY. Implementing non-thermal leptogenesis in this case is not 
so obvious. Finally, the axion models we have proposed appear 
to be compatible with natural SUSY where the $\mu$ parameter 
has magnitude of order few hundred GeV \cite{baer}. 

\acknowledgments{One of us (Q.S.) is supported in part by the 
DOE grant DOE-SC0013880. We would like to thank Vedat Nefer 
\c{S}eno\u{g}uz for useful discussions, and also for sharing 
with us his elegant analysis (unpublished) of axion physics 
and supersymmetric hybrid inflation in another model.}

\def\ijmp#1#2#3{{Int. Jour. Mod. Phys.}
{\bf #1},~#3~(#2)}
\def\plb#1#2#3{{Phys. Lett. B }{\bf #1},~#3~(#2)}
\def\zpc#1#2#3{{Z. Phys. C }{\bf #1},~#3~(#2)}
\def\prl#1#2#3{{Phys. Rev. Lett.}
{\bf #1},~#3~(#2)}
\def\rmp#1#2#3{{Rev. Mod. Phys.}
{\bf #1},~#3~(#2)}
\def\prep#1#2#3{{Phys. Rep. }{\bf #1},~#3~(#2)}
\def\prd#1#2#3{{Phys. Rev. D }{\bf #1},~#3~(#2)}
\def\npb#1#2#3{{Nucl. Phys. }{\bf B#1},~#3~(#2)}
\def\np#1#2#3{{Nucl. Phys. B }{\bf #1},~#3~(#2)}
\def\npps#1#2#3{{Nucl. Phys. B (Proc. Sup.)}
{\bf #1},~#3~(#2)}
\def\mpl#1#2#3{{Mod. Phys. Lett.}
{\bf #1},~#3~(#2)}
\def\arnps#1#2#3{{Annu. Rev. Nucl. Part. Sci.}
{\bf #1},~#3~(#2)}
\def\sjnp#1#2#3{{Sov. J. Nucl. Phys.}
{\bf #1},~#3~(#2)}
\def\jetp#1#2#3{{JETP Lett. }{\bf #1},~#3~(#2)}
\def\app#1#2#3{{Acta Phys. Polon.}
{\bf #1},~#3~(#2)}
\def\rnc#1#2#3{{Riv. Nuovo Cim.}
{\bf #1},~#3~(#2)}
\def\ap#1#2#3{{Ann. Phys. }{\bf #1},~#3~(#2)}
\def\ptp#1#2#3{{Prog. Theor. Phys.}
{\bf #1},~#3~(#2)}
\def\apjl#1#2#3{{Astrophys. J. Lett.}
{\bf #1},~#3~(#2)}
\def\apjs#1#2#3{{Astrophys. J. Suppl.}
{\bf #1},~#3~(#2)}
\def\n#1#2#3{{Nature }{\bf #1},~#3~(#2)}
\def\apj#1#2#3{{Astrophys. J.}
{\bf #1},~#3~(#2)}
\def\anj#1#2#3{{Astron. J. }{\bf #1},~#3~(#2)}
\def\mnras#1#2#3{{MNRAS }{\bf #1},~#3~(#2)}
\def\grg#1#2#3{{Gen. Rel. Grav.}
{\bf #1},~#3~(#2)}
\def\s#1#2#3{{Science }{\bf #1},~#3~(#2)}
\def\baas#1#2#3{{Bull. Am. Astron. Soc.}
{\bf #1},~#3~(#2)}
\def\ibid#1#2#3{{\it ibid. }{\bf #1},~#3~(#2)}
\def\cpc#1#2#3{{Comput. Phys. Commun.}
{\bf #1},~#3~(#2)}
\def\astp#1#2#3{{Astropart. Phys.}
{\bf #1},~#3~(#2)}
\def\epjc#1#2#3{{Eur. Phys. J. C}
{\bf #1},~#3~(#2)}
\def\nima#1#2#3{{Nucl. Instrum. Meth. A}
{\bf #1},~#3~(#2)}
\def\jhep#1#2#3{{J. High Energy Phys.}
{\bf #1},~#3~(#2)}
\def\jcap#1#2#3{{J. Cosmol. Astropart. Phys.}
{\bf #1},~#3~(#2)}
\def\lnp#1#2#3{{Lect. Notes Phys.}
{\bf #1},~#3~(#2)}
\def\jpcs#1#2#3{{J. Phys. Conf. Ser.}
{\bf #1},~#3~(#2)}
\def\aap#1#2#3{{Astron. Astrophys.}
{\bf #1},~#3~(#2)}
\def\mpla#1#2#3{{Mod. Phys. Lett. A}
{\bf #1},~#3~(#2)}


\begin{thebibliography}{}

\bibitem{dvali}
G.~Dvali, R.~Schaefer, and Q.~Shafi, Phys. Rev. Lett. 
{\bf 73}, 1886 (1994).
%%CITATION = doi:10.1103/PhysRevLett.73.1886;%%

\bibitem{copeland}
E.J.~Copeland, A.R.~Liddle, D.H.~Lyth, E.D.~Stewart, 
and D.~Wands, Phys. Rev. D {\bf 49}, 6410 (1994).
%%CITATION = doi:10.1103/PhysRevD.49.6410;%%

\bibitem{lectures}
G.~Lazarides, NATO Sci. Ser. II {\bf 34}, 399 (2001)
[hep-ph/0011130]; 
%%CITATION = HEP-PH/0011130;%%
\lnp{592}{2002}{351} [hep-ph/0111328];
%%CITATION = HEP-PH 0111328;%%
\jpcs{53}{2006}{528} [hep-ph/0607032].
%%CITATION = 00462,53,528;%%

\bibitem{planckcosm}
P.A.R.~Ade et al. [Planck Collaboration], 
Astron. Astrophys. {\bf 594}, A13 (2016). 
%%CITATION = doi:10.1051/0004-6361/201525830;%%

\bibitem{rehman}
M.U.~Rehman, Q.~Shafi, and J.R.~Wickman, Phys. Lett. 
B {\bf 683}, 191 (2010);
%%CITATION = doi:10.1016/j.physletb.2009.12.010;%%
C.~Pallis and Q.~Shafi, Phys. Lett. B {\bf 725}, 327 
(2013);
%%CITATION = doi:10.1016/j.physletb.2013.07.029;%%
W.~Buchm\"{u}ller, V.~Domcke, K.~Kamada, and K.~Schmitz, 
J. Cosmol. Astropart. Phys. {\bf 07}, 054 (2014).
%%CITATION = doi:10.1088/1475-7516/2014/07/054;%%

\bibitem{gravitywaves}
M.U.~Rehman, Q.~Shafi, and J.R.~Wickman, Phys. Rev. D 
{\bf 83}, 067304 (2011); 
%%CITATION = doi:10.1103/PhysRevD.83.067304;%%
M.~Civiletti, C.~Pallis, and Q.~Shafi, Phys. Lett. B 
{\bf 733}, 276 (2014).
%%CITATION = doi:10.1016/j.physletb.2014.04.060;%%

\bibitem{bastero}
M.~Bastero-Gil, S.F.~King, and Q.~Shafi, Phys. Lett. 
B {\bf 651}, 345 (2007).
%%CITATION = doi:10.1016/j.physletb.2006.06.085;%%

\bibitem{initial}
C.~Panagiotakopoulos and N.~Tetradis, \prd{59}{1999}{083502}; 
G.~Lazarides and N.~Tetradis, \prd{58}{1998}{123502}.
%%CITATION = HEP-PH/9802242;%% 
%%CITATION = HEP-PH/9710526;%%

\bibitem{PecceiQuinn}
R.D.~Peccei and H.R.~Quinn, Phys. Rev. Lett. 
{\bf 38}, 1440 (1977);
%%CITATION = doi:10.1103/PhysRevLett.38.1440;%%
S.~Weinberg, Phys. Rev. Lett. {\bf 40}, 223 (1978); 
%%CITATION = doi:10.1103/PhysRevLett.40.223;%%
F.~Wilczek, Phys. Rev. Lett. {\bf 40}, 279 (1978).
%%CITATION = doi:10.1103/PhysRevLett.40.279;%%

\bibitem{preskill}
J.~Preskill, M.B.~Wise, and F.~Wilczek, 
Phys. Lett. {\bf 120B}, 127 (1983); 
%%CITATION = doi:10.1016/0370-2693(83)90637-8;%%
L.F.~Abbott and P.~Sikivie, Phys. Lett. {\bf 120B}, 
133 (1983); 
%%CITATION = doi:10.1016/0370-2693(83)90638-X;%%
M.~Dine and W.~Fischler, Phys. Lett. {\bf 120B}, 137 
(1983).
%%CITATION = doi:10.1016/0370-2693(83)90639-1;%%

\bibitem{nilles}
E.J.~Chun, J.E.~Kim, and H.P.~Nilles, Nucl. Phys. B 
{\bf 370}, 105 (1992);
%%CITATION = doi:10.1016/0550-3213(92)90346-D;%%
G.~Lazarides and Q.~Shafi, Phys. Rev. D {\bf 58}, 
071702 (1998).
%%CITATION = doi:10.1103/PhysRevD.58.071702;%%

\bibitem{dinf}
G.~Lazarides and C.~Panagiotakopoulos, Phys. Rev. D 
{\bf 92}, 123502 (2015).
%%CITATION = doi:10.1103/PhysRevD.92.123502;%%

\bibitem{baer}
H.~Baer, AIP Conf. Proc. {\bf 1743}, 050002 (2016)
[arXiv: 1510.07501] and references therein.

\bibitem{sikivie} 	
P.~Sikivie, Phys. Rev. Lett. {\bf 48}, 1156 (1982).
%%CITATION = doi:10.1103/PhysRevLett.48.1156;%%

\bibitem{yanagida}
K.~Harigaya, M.~Ibe, M.~Kawasaki, and T.T. Yanagida, 
J. Cosmol. Astropart. Phys. {\bf 11}, 003 (2015).
%%CITATION = doi:10.1088/1475-7516/2015/11/003;%%

\bibitem{intermediate}
G.~Lazarides and Q.~Shafi, Phys. Lett. B {\bf 489}, 194 
(2000).
%%CITATION = doi:10.1016/S0370-2693(00)00904-7;%%

\bibitem{freeze}
K.~Griest, M.~Kamionkowski, and M.S.~Turner, Phys. Rev. 
D {\bf 41}, 3565 (1990).
%%CITATION = doi:10.1103/PhysRevD.41.3565;%%

\bibitem{neutralino} 
G.~Lazarides, Lect. Notes Phys. {\bf 720}, 3 (2007) 
[hep-ph/0601016].
%%CITATION = doi:10.1007/978-3-540-71013-4_1;%%

\bibitem{saxiondensity}
M. Kawasaki and K. Nakayama, Ann. Rev. Nucl. Part. Sci. 
{\bf 63}, 69 (2013).
%%CITATION = doi:10.1146/annurev-nucl-102212-170536;%%

\bibitem{axinodensity}
E.J.~ Chun, Phys. Rev. D {\bf 84}, 043509 (2011); 
%%CITATION = doi:10.1103/PhysRevD.84.043509;%%
K.J.~Bae, K.~Choi, and S.H.~Im, J. High Energy Phys. 
{\bf 08}, 065 (2011);
%%CITATION = doi:10.1007/JHEP08(2011)065;%%
K.J.~Bae, E.J.~Chun, and S.H.~Im, \jcap{03}{2012}{013}.
%%CITATION = doi:10.1088/1475-7516/2012/03/013;%%

\bibitem{thermalaxions}
P.~Graf and F.D.~ Steffen, Phys. Rev. D {\bf 83}, 075011 
(2011); 
%%CITATION = doi:10.1103/PhysRevD.83.075011;%%
A.~Salvio, A.~Strumia, and W.~Xue, \jcap{01}{2014}{011}.
%%CITATION = doi:10.1088/1475-7516/2014/01/011;%%

\bibitem{lyth}
D.H.~Lyth, Phys. Rev. D {\bf 45}, 3394 (1992).
%%CITATION = doi:10.1103/PhysRevD.45.3394;%%

\bibitem{planck}
P.A.R.~Ade et al. [Planck Collaboration], 
Astron. Astrophys. {\bf 594}, A20 (2016). 
%%CITATION = doi:10.1051/0004-6361/201525898;%%

\bibitem{gondolo}
L.~Visinelli and P.~Gondolo, Phys. Rev. D {\bf 80}, 
035024 (2009).
%%CITATION = doi:10.1103/PhysRevD.80.035024;%%

\bibitem{curvaton}
K. Dimopoulos, G. Lazarides, D. Lyth, and R. Ruiz de Austri, 
\jhep{05}{2003}{057}.
%%CITATION = doi:10.1088/1126-6708/2003/05/057;%%

\bibitem{axionwalls}
G.~Lazarides and Q.~Shafi, Phys. Lett. {\bf 115B}, 21 (1982);
%%CITATION = doi:10.1016/0370-2693(82)90506-8;%%
H.~Georgi and M.B.~Wise, Phys. Lett. {\bf 116B}, 123 (1982).
%%CITATION = doi:10.1016/0370-2693(82)90989-3;%%

\bibitem{cavity}
L.~Duffy, P.~Sikivie, D.B.~Tanner, S.J.~Asztalos, C.~Hagmann, 
D.~Kinion, L.J.~ Rosenberg, K.~van Bibber, D.~Yu, and 
R.F.~Bradley, Phys. Rev. Lett.{\bf 95}, 091304 (2005). 
%%CITATION = doi:10.1103/PhysRevLett.95.091304;%%

\bibitem{trotta}
G.~Lazarides, R.~Ruiz de Austri, and R.~Trotta, Phys. Rev. 
D {\bf 70}, 123527 (2004). 
%%CITATION = doi:10.1103/PhysRevD.70.123527;%%

\bibitem{magg}
G.~Lazarides, M.~Magg, and Q.~Shafi, Phys. Lett. {\bf 97B}, 87 
(1980).
%%CITATION = doi:10.1016/0370-2693(80)90553-5;%%

\bibitem{shift}
R.~Jeannerot, S.~Khalil, G.~Lazarides, and Q.~Shafi, 
J. High Energy Phys. {\bf 10}, 012 (2000);
%%CITATION = doi:10.1088/1126-6708/2000/10/012;%%
R.~Jeannerot, S.~Khalil, and G.~Lazarides, J. High Energy Phys. 
{\bf 07}, 069 (2002).
%%CITATION = doi:10.1088/1126-6708/2002/07/069;%%

\bibitem{smooth}
G.~Lazarides and C.~Panagiotakopoulos, Phys. Rev. D {\bf 52}, 
R559 (1995);
%%CITATION = doi:10.1103/PhysRevD.52.R559;%%
M.U.~Rehman, V.N.~\c{S}eno\u{g}uz, and Q.~Shafi, Phys. Rev. D 
{\bf 75}, 043522 (2007);
%%CITATION = doi:10.1103/PhysRevD.75.043522;%%
G.~Lazarides and A.~Vamvasakis, Phys. Rev. D {\bf 76}, 083507 
(2007).
%%CITATION = doi:10.1103/PhysRevD.76.083507;%%

\bibitem{thermallepto} 
M.~Fukugita and T.~Yanagida, Phys. Lett. B {\bf 174}, 45 (1986).
%%CITATION = doi:10.1016/0370-2693(86)91126-3;%%

\bibitem{nonthermalepto}
G.~Lazarides and Q.~Shafi, Phys. Lett. B {\bf 258}, 305 (1991);
%%CITATION = PHLTA,B258,305;%%
G.~Lazarides, R.K.~Schaefer, and Q.~Shafi, Phys. Rev. D {\bf 56}, 
1324 (1997);
%%CITATION = doi:10.1103/PhysRevD.56.1324;%%
G.~Lazarides, Q.~Shafi, and N.D.~Vlachos, Phys. Lett. B {\bf 427}, 
53 (1998).
%%CITATION = doi:10.1016/S0370-2693(98)00306-2;%%
 
\bibitem{dvali1}
G.R.~Dvali, G.~Lazarides, and Q.~Shafi, Phys. Lett. B 
{\bf 424}, 259 (1998). 
%%CITATION = doi:10.1016/S0370-2693(98)00145-2;%%

\bibitem{gravitino}
M.Yu.~Khlopov and A.D.~Linde,
Phys. Lett. {\bf 138B}, 265 (1984);
%%CITATION = PHLTA,B138,265;%%
J.~Ellis, J.E.~Kim, and D.~Nanopoulos,
Phys. Lett. {\bf 145B}, 181 (1984);
%%CITATION = PHLTA,B145,181;%%
J.R.~Ellis, D.V.~Nanopoulos, and S.~Sarkar,
\npb{259}{1985}{175}.
%%CITATION = NUPHA,B259,175;%%

\bibitem{split}
N.~Okada and Q.~Shafi, PoS PLANCK2015, 121 (2015) 
[arXiv:1506.01410].
%%CITATION = ARXIV:1506.01410;%%

\end{thebibliography}
\end{document}